\documentclass[journal=jacsat,manuscript=article]{achemso}
\usepackage{mathrsfs}
\usepackage{amsmath,textcomp,gensymb}
\usepackage{amsfonts}
\usepackage{amssymb}
\usepackage{amsthm}
\usepackage{graphicx}
\usepackage{natbib}
\usepackage{color}
\usepackage{hyperref}
\usepackage{bm}
\usepackage[caption=false]{subfig}
\usepackage{verbatim}
\usepackage{soul,ulem}
\usepackage[version=3]{mhchem}

\author{Mirko Rocci}
\email{rocci@mit.edu}
\affiliation{Francis Bitter Magnet Laboratory \& Plasma Science and Fusion Center\\Massachusetts Institute of Technology, Cambridge, MA-02139, USA}
\alsoaffiliation{NEST, Instituto Nanoscienze-CNR and Scuola Normale Superiore, I-56127 Pisa, Italy}
\altaffiliation{Contributed equally to this work}
\author{Dhavala Suri}
\email{dmsuri@mit.edu}
\affiliation{Francis Bitter Magnet Laboratory \& Plasma Science and Fusion Center\\Massachusetts Institute of Technology, Cambridge, MA-02139, USA}
\altaffiliation{Contributed equally to this work}
\author{Akashdeep Kamra}
\affiliation{Center for Quantum Spintronics, Department of Physics, Norwegian University of Science and Technology, NO-7491 Trondheim, Norway}
\author{Gilv$\hat{\text{a}}$nia Vilela}
\affiliation{Francis Bitter Magnet Laboratory \& Plasma Science and Fusion Center\\Massachusetts Institute of Technology, Cambridge, MA-02139, USA}
\alsoaffiliation{Física de Materiais, Universidade de Pernambuco, Recife, 50720-001, Brazil}
\author{Yota Takamura}
\affiliation{School of Engineering, Tokyo Institute of Technology, Tokyo 152-8550, Japan}
\author{Norbert M. Nemes}
\affiliation{GFMC, Departamento de Física de Materiales, Universidad Complutense de Madrid, E-28040 Madrid, Spain}
\author{Jose L. Martinez}
\affiliation{Instituto de Ciencia de Materiales de Madrid, C.S.I.C., Cantoblanco, E-28049 Madrid, Spain}
\author{Mar Garcia Hernandez}
\affiliation{Instituto de Ciencia de Materiales de Madrid, C.S.I.C., Cantoblanco, E-28049 Madrid, Spain}
\author{Jagadeesh S. Moodera}
\email{moodera@mit.edu}
\affiliation{Francis Bitter Magnet Laboratory \& Plasma Science and Fusion Center\\Massachusetts Institute of Technology, Cambridge, MA-02139, USA}
\alsoaffiliation{Department of Physics, Massachusetts Institute of Technology, Cambridge, MA-02139, USA}
	
\title{Large Enhancement of Critical Current in Superconducting Devices by Gate Voltage}

\begin{document}

\begin{abstract}  
Significant control over the properties of a high-carrier density superconductor via an applied electric field has been considered infeasible due to  screening of the field over atomic length scales. Here, we demonstrate an enhancement of up to 30~\% in critical current in a back-gate tunable NbN, micro- and nano superconducting bridges. Our suggested plausible mechanism of this enhancement in critical current based on surface nucleation and pinning of Abrikosov vortices is consistent with expectations and observations for type-II superconductor films with thicknesses comparable to their coherence length. Furthermore, we demonstrate an applied electric field dependent infinite electroresistance and hysteretic resistance.  Our work  presents an electric field driven enhancement in the superconducting property in type-II superconductors which is a crucial step towards the  understanding of  field-effects on the fundamental properties of a superconductor and its exploitation for logic and memory applications in a superconductor-based low-dissipation digital computing paradigm. 
\end{abstract}


\section{Introduction}Semiconductor-based field-effect transistors (FETs), which have been instrumental in the silicon revolution, operate through modulation of resistance between the source and drain electrodes via an applied gate voltage. This modulation, in turn, is achieved via a change in the charge carrier density resulting from the electric field generated by the gate voltage. The relatively low carrier densities in semiconductors allows for a strong resistance modulation with reasonable gate voltages thereby enabling broad functionalities. Such a field effect is not expected to work with metals, which have a very {\it high} charge density compared to what can be induced by a gate voltage, and was shown to be negligibly weak ~\cite{Bonfiglioli1956,Bonfiglioli1959}.
	
	Gate-voltage modulation of superconductivity is considered immensely useful, and has been attempted for some time~\cite{glover,Ahn2003,Xi1992,Frey1995}, including the recent advances in gate tunability of superconductivity in Van der Waals materials \cite{Sajadi922,Costanzo2016,Hanzawa3986}. Conventional superconductors have been amenable to control via interaction with magnetic fields~\cite{wangj,Giazotto2015,Linder2015,Moraru2006,Khaire2009,Bergeret2018}, but not electric fields. Superconducting properties, such as critical temperature $T_c$  of metallic superconductors, were found to be fairly insensitive to gate voltages~\cite{glover,Lipavsky2010}, exhibiting a minuscule change of $\sim 10^{-3}\%$. In contrast, unconventional superconductors based on strongly correlated oxides allow for an efficient gate-modulation due to their relatively {\it low} carrier concentration~\cite{Xi1992,Frey1995,Ahn2003}. The change in density of states at the chemical potential, which is associated with the gate-modulated carrier density, alters the superconducting order parameter and qualitatively explains the experimental observations discussed above~\cite{Frey1995,Lipavsky2010}. This implies that superconducting properties, such as $T_c$, can be enhanced (reduced) by an increase (decrease) in the carrier concentration via a positive (negative) gate voltage. The change in superconducting properties is thus odd in the gate voltage, i.e. {\it unipolar}. Besides this quasi-equilibrium modulation, various nonequilibrium approaches to controlling the superconducting properties using electromagnetic radiation have been successful, with significant advances demonstrated recently~\cite{Cavalleri2018,Demsar2020}.

	In contrast with previous literature and expectations~\cite{LONDON1935341,Lipavsky2010}, De Simoni and coworkers recently reported a gate-voltage-induced suppression of the critical current ($I_c$) in all-metallic superconducting nano-bridges and junctions~\cite{DeSimoni2018,desimoni2019josephson,simoni2020niobium,puglia}. Furthermore, the observed suppression is even in the gate voltage, i.e. {\it bipolar}. Apart from the technological potential, these observations have raised two fundamental questions regarding (i) how a gate-voltage-induced electric field can affect a high-carrier density superconductor~\cite{LONDON1935341,PIATTI201817,Lipavsky2010,glover,2006.07091}, and (ii) what mechanism causes a change in the $I_c$. These crucial issues remain unaddressed thus far, although the possibility of metallic puddles creation~\cite{paolo}, that could reduce the $I_c$, has been floated. Various other mechanisms that may degrade superconductivity could be envisaged as accounting for the  observed $I_c$ reduction; for instance injection of high-energetic carriers as established by two most recent experiments with compelling evidences considering the size of leakage currents  \cite{2005.00584,2005.00462}.
	
	\begin{figure}[tbh]
		\centering
		\includegraphics[width=14cm]{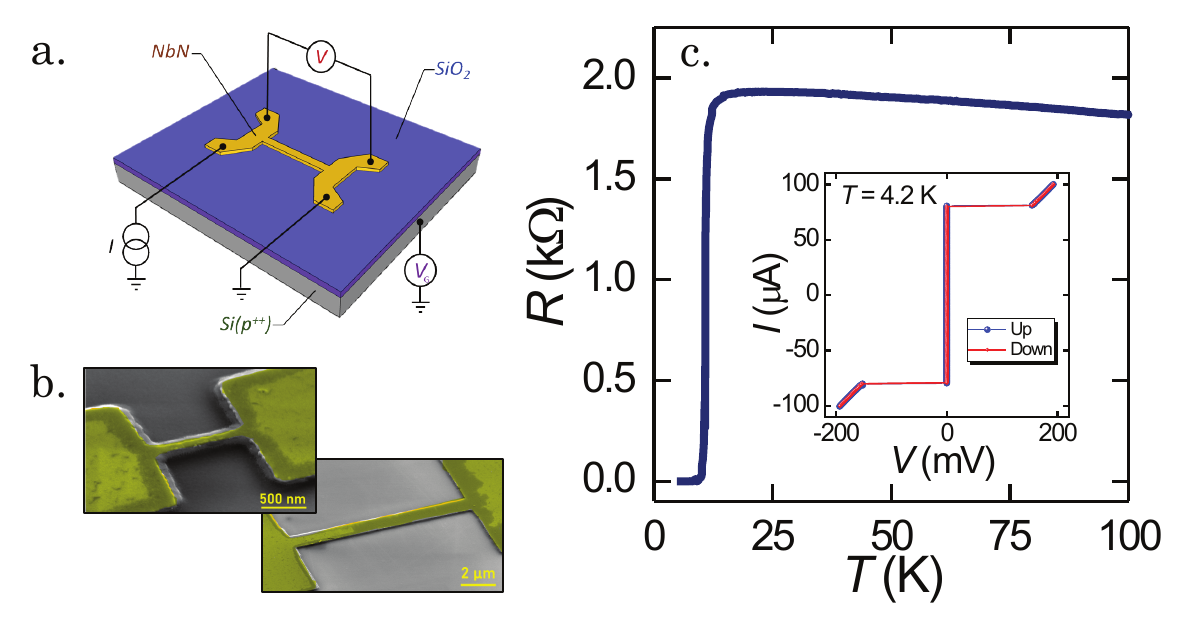}
		\caption{NbN device schematics and characterization. (a) Schematic depicting the measurement geometry of NbN bridges. (b) Pseudo-color scanning electron micrographs [image taken with sample tilted by 55$^{\circ}$] of the fabricated 100 nm wide NB [left] and 1 $\mu$m wide $\mu$B [right]. (c) Resistance versus temperature variation for the $\mu$B device showing the superconducting transition at $\approx$ 10.8 K. Inset shows current-voltage characteristics of the device at 4.2 K (for both forward and backward scans).}
		\label{fig:1} 
	\end{figure}
	
	Here, we demonstrate a {\it bipolar} gate-voltage-induced {\it enhancement} in the $I_c$ of NbN (high carrier density superconductor) based superconducting bridges by  30$\%$, while the critical temperature remains insensitive to the applied voltage. Besides uncovering novel fundamental phenomena, we demonstrate infinite electroresistance, i.e. gate-voltage-controlled change in resistance between zero and a finite value, and hysteretic resistance variation vs. gate voltage. These two effects could be exploited for low-dissipation logic and memory elements based on superconductors. We further demonstrate that the observed phenomena work for bridges in the nanoscale providing a proof-of-principle for scalability of such a technological paradigm. We also qualitatively discuss a plausible mechanism for the observed critical current modulation with the gate voltage. Hypothesizing that the critical current in our films is the value at which flux creep from one edge to the other becomes energetically favorable~\cite{Shmidt1970A,Shmidt1970B} by overcoming Bean-Livingston surface barrier~\cite{Bean1964}, a gate-voltage-induced enhancement of this surface barrier could account for our experiments and is consistent with related literature~\cite{Shmidt1970A,Shmidt1970B,Aslamazov1975,Mawatari1994,Ilin2014,Clem2011}. Altogether, our work provides crucial insights for understanding the field effect in metallic superconductors demonstrating that it could be further optimized with suitable surface termination and employed for enhancing, instead of suppressing, superconducting properties.
	 
	\section{Synthesis and Fabrication} Niobium nitride (NbN) thin films with thicknesses $t = 10 $ nm and $7$ nm were grown on Si/SiO$_2$ substrates (Fig.~\ref{fig:1}). The 300 nm thick SiO$_2$ layer ensured electrical isolation between the superconducting film and the \textit{p}-doped Si substrate acting as the gate. NbN thin films with Al$_2$O$_3$ ($t = 5$ nm) capping were grown in situ by reactive DC magnetron sputtering (for NbN) and by standard RF non-reactive magnetron sputtering (for Al$_2$O$_3$). Substrates were annealed at 573~K for 1 hour in UHV prior to the deposition. The base pressure of the sputtering chamber before the film deposition was below 5 $\times$ 10$^{-8}$ Torr. The thin film growth was found to be poly-crystalline in texture with sharp, well defined NbN characteristic peaks in x-ray diffraction spectrum (see supplementary information). The multilayer structures were then patterned via e-beam lithography into a microbridge ($\mu$B), with length $l = 10~\mu$m, width $w = 1~\mu$m, and thickness $t = 10~$nm, and a nanobridge (NB) with $l = 1~\mu$m, $w = 100~$nm, and $t = 7~$nm (Fig.~\ref{fig:1}). Negative-tone resist was spun on the film and exposed at 10 kV following a soft bake. The resulting pattern was developed and the samples were then Ar$^+$ ion-milled  to fabricate the bridges.

\section{Results}We first present our experiments on the $\mu$B device. It was cooled down below its transition temperature into the superconducting state. A transition temperature T$_c$ $\approx$ 10.8 K can be seen in the temperature dependence of resistance shown in Fig. \ref{fig:1}(c),  similar to previous NbN thin films, (for example -- ref.\cite{Romestain2004}). The inset depicts the corresponding I-V characteristics at 4.2 K showing a $I_c$ of about 82.5 $\mu$A. The Bardeen-Cooper-Schrieffer (BCS) energy gap 2$\Delta_0$ = 4.05~k$_B$T$_c$ corresponds to~$\approx$~4.16~meV, where k$_B$ is Boltzmann constant \cite{tinkham,KOMENOU1968335}. The London penetration depth is derived from the above parameters as $\lambda_L = \sqrt{\hslash R_N w t/\pi \mu_0l\lambda_0} \approx 450$~nm, where $ l = 10~\mu$m, $w = 1~\mu$m, and $t = 10$~nm are length, width, and thickness of the bridge, respectively. $R_N$ = 2~k$\Omega$ is the resistance in the normal state at low temperature, and $\mu_0$ is magnetic permeability in vacuum. The Ginzburg-Landau coherence length is estimated as $\xi_{GL} = \sqrt{\hslash l/R_{N}wt N_{F}e^2\Delta_0}~\approx~9 $~nm, where $N_F = 1.65 \times 10^{28}/(m^{3}eV)$ is the density of states in NbN at Fermi level \cite{PhysRevB.77.214503} and $e$ is the electronic charge. This places the Pearl length~\cite{Pearl1964,Romijn1982} ($2 \lambda_L^2 / t$) at around 40 $\mu$m ($\gg w$ ) ensuring a spatially uniform current through the film.
	
	\begin{figure}[tb]
		\centering
		\includegraphics[width=10cm]{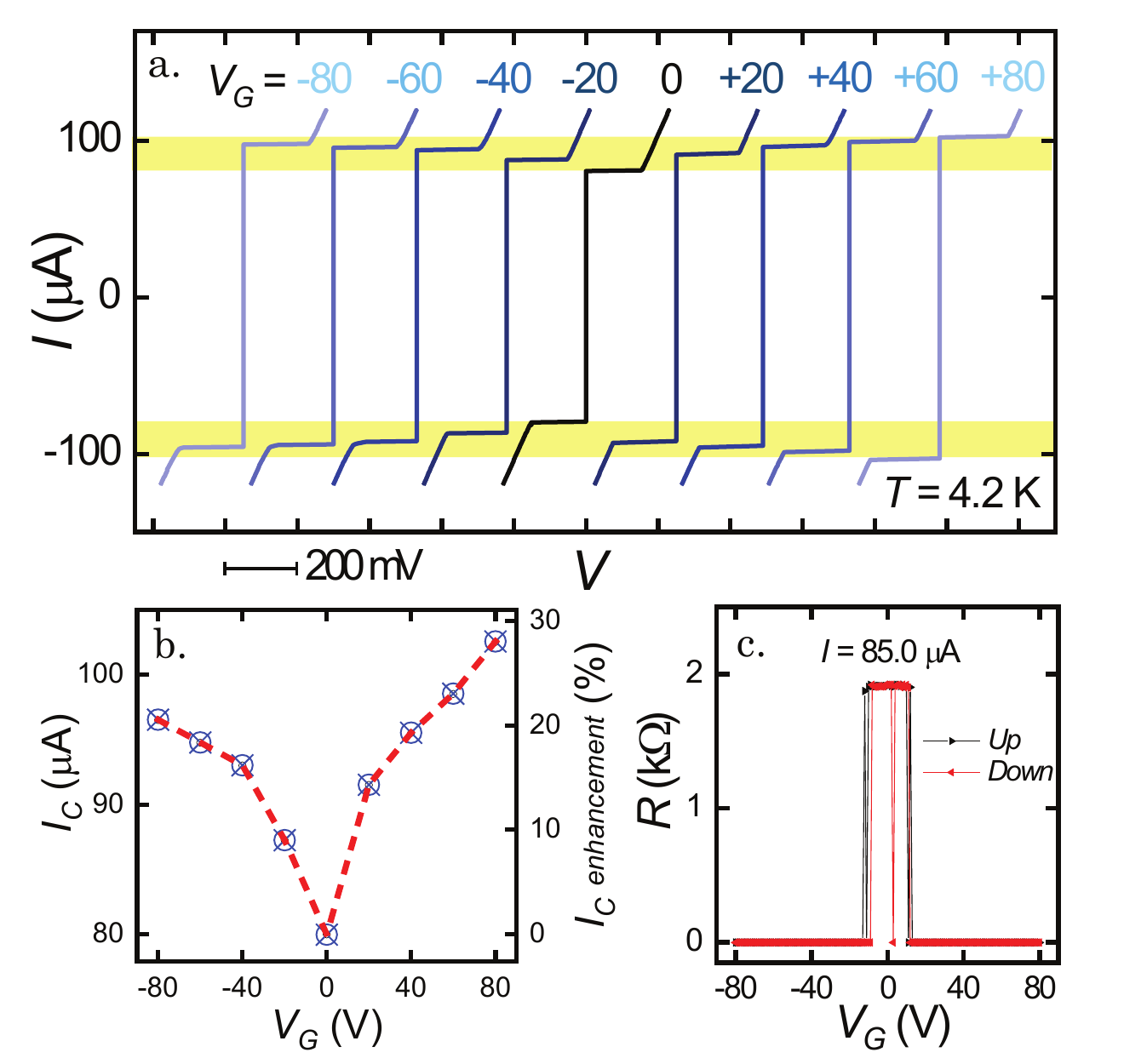}
		\caption{Gate-voltage-control of the $\mu$B properties at 4.2 K. (a) Current-voltage characteristics for various values of the back-gate voltage V$_G$. (b) $I_c$ as a function of back-gate voltage. Scale on the right shows the corresponding enhancement percentage. (c) Resistance of $\mu$B as a function of back gate voltage at a constant current bias of 85 $\mu$A.}
		\label{fig:2}
	\end{figure}
	
	Further, the same $\mu$B was investigated for its electric field response by applying back-gate voltage as depicted in Fig.~\ref{fig:1}(a). Figure \ref{fig:2} (a) shows current-voltage characteristics of the $\mu$B at different back-gate voltages ($V_G$ varying from -80 V to +80 V) at 4.2~K. The $I_c$ enhances with increase in gate voltage from $\approx$ 80 $\mu$A to 105 $\mu$A (i.e., an increase in critical current density, $J_c$ from 0.80~$\times~10^6$~Acm$^{-2}$ to 1.03~$\times~10^6$~Acm$^{-2}$).  The enhancement shows a nearly symmetric response with respect to the gate voltage polarity [Fig.~\ref{fig:2} (b)] and is observed to be  $\approx$ 30\% which is the largest modulation to date \cite{DeSimoni2018}. The applied voltage, however, did not produce any observable change in the critical temperature. As discussed below, the observed {\it enhancement} in the $I_c$, as compared to the previously reported suppression~\cite{DeSimoni2018}, may be attributed to our choice of superconductor (type II) and the film thickness, that is comparable to the superconducting coherence length.
	
	Next, we examine the resistance variation with gate voltage [Fig. \ref{fig:2}(c)] biasing the device at a constant current of 85 $\mu$A. The device completely recovers the superconducting state from normal state for a finite value of gate voltage $<$  20~V. This modulation of resistance may be used to define an ``electroresistance'' $ER$, similar to the well-known magnetoresistance~\cite{Moodera1995,Moodera1999,Zutic2004,Fert2008}, $ER \equiv (R_{max}-R_{min})/R_{min}$, which yields infinite value for our $\mu$B. The response is symmetric with respect to gate voltage polarity and is a direct consequence of the gate-voltage-induced $I_c$ enhancement. We rule out the possibility of such a response as being due to heating or electronic refrigeration effects as mentioned in Ref.~\cite{RevModPhys.78.217} by noting that the measurements were performed by immersing the sample in liquid Helium in a storage dewar. This helps maintain the sample in thermodynamic equilibrium. We also remark that the leakage current due to the applied electric field was $\leq$ 1~nA. Any spurious effect due to gate leakage can be ruled out as it would only lead to a reduction in critical current.
	
	\begin{figure}[tb]
		\centering
		\includegraphics[width=10 cm]{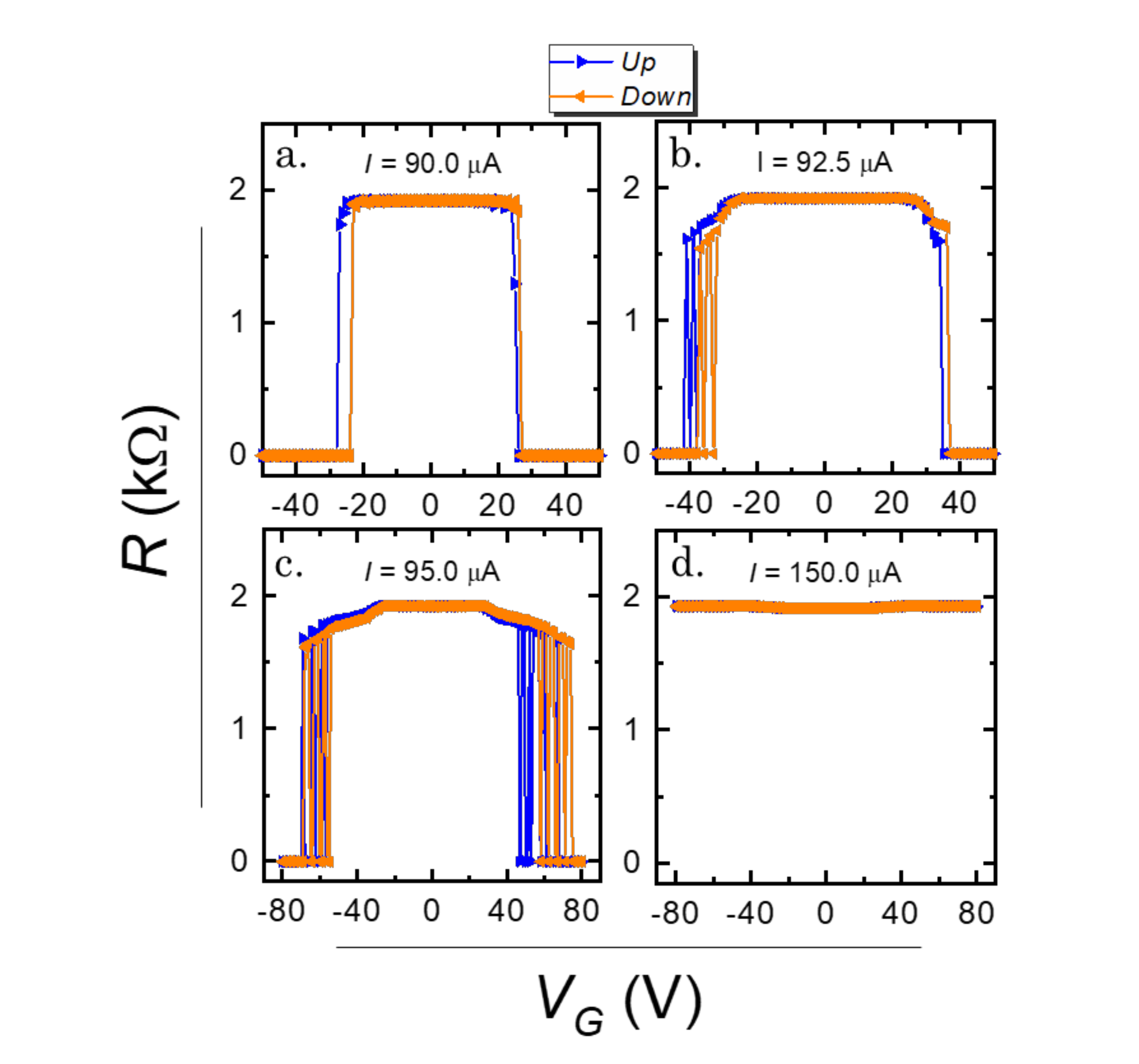}
		\caption{Back-gate-controlled switching between superconducting and normal states at different bias currents at 4.2~K. Resistance of the $\mu$B as a function of back-gate voltage for bias currents of (a) 90 $\mu$A, (b) 92.5 $\mu$A, (c) 95 $\mu$A, and (d) 150~$\mu$A. Both forward and backward sweeps are shown.}
		\label{fig:3}
	\end{figure}

	We also study electric field-induced switching of the device by applying different bias currents (90 $\mu$A, 92.5 $\mu$A, 95 $\mu$A and 150 $\mu$A) as shown in Fig.~\ref{fig:3}. The back-gate voltage was scanned for both upward (negative to positive) and downward (positive to negative) directions. While we are successfully and consistently, able to drive the system from superconducting to normal state and vice versa, we observe slight hysteresis with gate voltage sweeps as it approaches the transition voltage. With increase in the bias current, the gate voltage required for the transition is higher and the hysteresis becomes more prominent. This could arise as a consequence of charge pinning due to surface inhomogeneities in the thin film. The range over which the quasi-normal state exists broadens with increase in the bias current which may be attributed to inhomogeneous superconducting state at higher currents or intrinsic thermal excitation in the sample, and not due to phase dynamics in the superconductor. Such scaling of the area under the hysteresis curve with bias current makes our devices a potential candidate for cryogenic memory systems~\cite{DeSimoni2018,Gingrich2016}. However the hysteresis may weaken in thinner films~\cite{PhysRevB68134515}. When the bias current is set to a relatively large value of 150 $\mu$A, the system does not achieve the superconducting state [Fig. \ref{fig:3}(d)] within the limits of gate voltage allowed by the SiO$_2$ dielectric. 
	
	\begin{figure}[tb]
		\includegraphics[width=16cm]{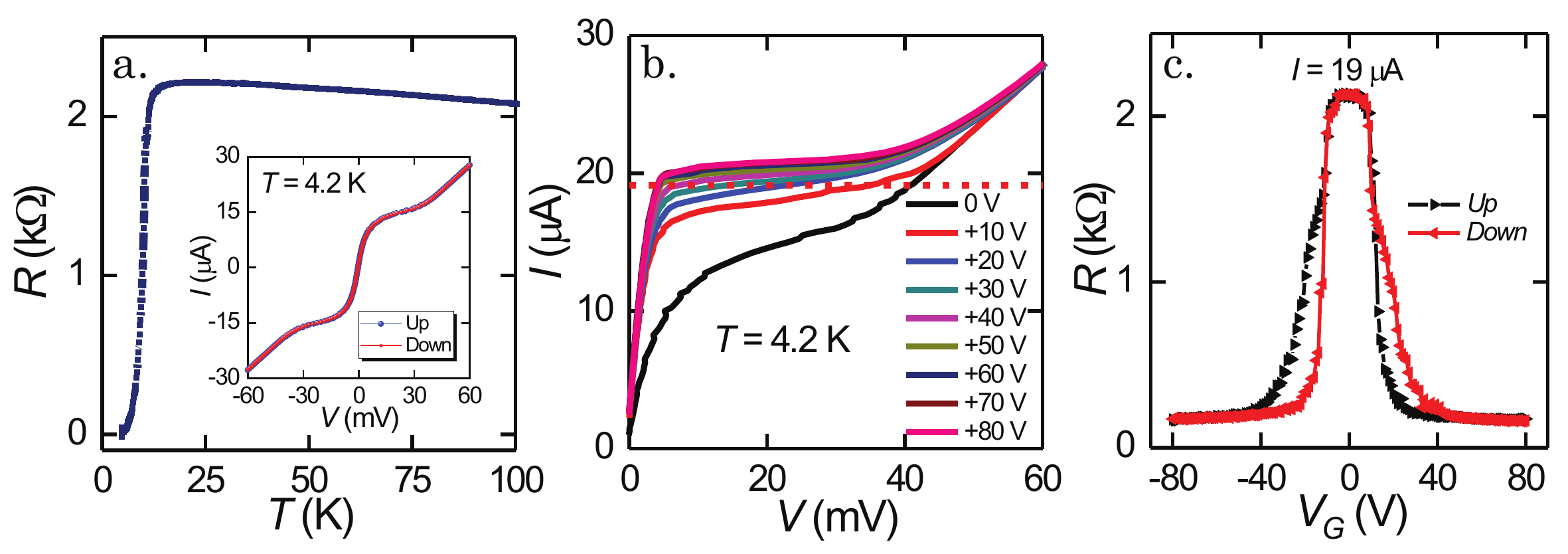}
		\begin{centering}   
		\caption{Gate-voltage-control of the NB at 4.2~K. (a) Resistance versus temperature. Inset shows the I-V characteristics without any gate bias (both forward and backward scans). (b) I-V characteristics at different gate voltages (c) Resistance versus $V_G$ at a fixed bias current of 19~$\mu$A. }
			\label{fig:4} \end{centering}
	\end{figure}
	
	In order to examine the dependence of gating effect on the bridge dimensions and probe the device scalability to nano-regime, we now present results for the NB device with thickness $t = 7~$nm ($t < \xi_{GL}$). Here, the aspect ratio $w/l = 1/10$ was kept the same as that of the $\mu$B. The resistance versus temperature curve in Fig.~\ref{fig:4}(a) shows a superconducting transition close to 12~K. However, the \textit{I-V} characteristics show an additional normal metal behavior with finite resistance at 4.2~K [inset of Fig.~\ref{fig:4}(a)]. This appears to be the result of the edge disorder caused by Ar ion milling and the concomitant degradation of the nanowire causing a small drop in the $T_c$~\footnote{We expect a sharper \textit{I-V} characteristic at lower temperatures.}. Nevertheless, in these NBs we observe a transition in Ic-like feature with gate bias similar to that of the  $\mu$B device discussed previously. On biasing with a constant current of 10~$\mu$A and scanning the gate voltage, the NB recovers the superconducting state with a much broader hysteresis [Fig.~\ref{fig:4}(c)] possibly due to increased charge pinning effects. The $ER$ in this case is nearly 1400\%, which is extremely large but finite since the NB does not transition to a completely superconducting state at 4.2~K.  We further note the absence of re-trapping currents \cite{courtois2008origin} in all our devices, as seen via \textit{I-V} characteristics i.e., symmetric in both forward and backward sweeps [Fig.~\ref{fig:1}~(c) and Fig.~\ref{fig:4}~(a)]. This observation is in contrast with previous works  \cite{DeSimoni2018,desimoni2019josephson,paolo} and suggests the role of symmetric flux pinning edges.
	
	Our experiments on both $\mu$B and NB devices demonstrate a robust coupling of $I_c$ to gate-voltage exhibiting an infinite (large) $ER$ in $\mu$B (NB). Despite the $T_c$ drop in the NB on account of additional disorder, the qualitative effects reported herein remain the same as for the $\mu$B thereby demonstrating their scalability for technological applications. Furthermore, the larger hysteresis in the NB should be beneficial for cryogenic memory devices. The gate voltage required to control the superconductor-normal state transition can be brought down significantly, by engineering the oxide layer thickness, to values comparable with the contemporary silicon technology. Finally, we note that similar measurements, presented in the supplementary information on 5 different devices with the same aspect ratio find essentially the same effects as discussed above.
	
	
	{\it Mechanism and discussion.---} We now discuss a plausible mechanism for the observed enhancement in the $I_c$. Depending on the sample details, such as physical dimensions~\cite{Pearl1964,Romijn1982,Korneeva2018}, grains and so on, $I_c$ can be determined by a variety of processes such as critical pair-breaking~\cite{Bardeen1962,DeGennes:566105,Romijn1982}, superconducting weak links~\cite{Likharev1979}, an intrinsic proximity effect~\cite{Charaev2017}, and surface vortex (flux) nucleation and flow~\cite{Shmidt1970A,Shmidt1970B,Aslamazov1975}. For a homogeneous type II superconductor with $t$ larger than or comparable to $\xi$, it becomes energetically favorable for vortices nucleating at one edge to move across to the other at large enough currents. This instability of the vortex system then determines the $I_c$~\cite{Shmidt1970A,Shmidt1970B,Aslamazov1975,Likharev1979,Mawatari1994,Ilin2014,Clem2011} and has been the basis for understanding experiments including magnetic field induced enhancement in $I_c$ of bent superconductors~\cite{Ilin2014}. 
	
	Working under the assumption that the vortex instability mechanism discussed above determines the critical current $I_c$ in our films, an applied gate voltage can influence $I_c$ by changing the vortex surface barrier~\cite{Shmidt1970A,Bean1964}. This appears consistent with two key features. First, our experimental observation that although $I_c$ is affected, critical temperature remains unaltered. A change in vortex surface barrier should not alter critical temperature. Second, the expectation that electric field is screened over atomic length scales in our high-carrier density superconductor~\cite{Lipavsky2010}. The gate voltage, therefore, drops largely over the interfaces causing strong interfacial electric fields. The latter should give rise to an interfacial Rashba spin-orbit interaction (SOI) parameterized by $\alpha \propto V_{G}$. We hypothesize that the ensuing interfacial SOI contributes an amount $\sim \alpha^2 \propto V_{G}^2$ to the vortex surface barrier and discuss its origin further in the supplementary information. The vortex surface barrier, and thus the change in critical current, is therefore bipolar. In this manner, we are able to explain all the key features of our experiments with the caveat that our reasonable sounding hypotheses need to be examined via further theoretical and experimental analyses.
	 
	In our considerations above, we have assumed a homogeneous superconducting state consistent with existing experimental evidence for samples similar to ours~\cite{Romestain2004}.  Finally, a possible role of magnetic impurities in determining the electric-field-induced vortex pinning in our devices cannot be ruled out~\cite{PhysRevLett.118.077001}. As detailed in the supplementary information, we have also observed similar gating effects in NbN/GdN bilayer films, where the ferromagnetic GdN layer may play a role via exchange-coupling to the superconducting NbN layer.
	
\section{Conclusion}We have demonstrated gate-voltage-induced enhancement by up to $30\%$ in $I_c$ of NbN-based superconducting bridges. We have put forward a qualitative plausible model that explains our experiments in terms of gate-voltage-controlled surface pinning of vortices. Capitalizing on this voltage control, we demonstrate infinite electroresistance and hysteretic resistance variation in our devices making them promising candidates for logic and memory applications. Our work thus provides fundamental new insights into the field-effect in superconductors, paving the way for even larger voltage-controlled enhancement of superconducting properties, and developing novel low-dissipation computing paradigms.

\begin{acknowledgement}
    	Authors thank A. K. Saydjari, Prof. Jacobo Santamaria, Prof. Carlos Leon, Morten Amundsen, and Marius Kalleberg Hope for fruitful discussions. M.R. was supported by the Marie Skłodowska-Curie grant agreement EuSuper No. 796603 under the European Union's Horizon 2020 research and innovation program. D.S. was supported by CIQM-NSF DMR-1231319, NSF Grant DMR1700137 and ONR Grant N00014-16-1-2657. Research at MIT was supported by NSF Grant DMR1700137, ONR Grant N00014-16-1-2657 and ARO grant W911NF1920041. A.K. acknowledges financial support from the Research Council of Norway through its Centers of Excellence funding scheme, project 262633, ``QuSpin''. G.V. was supported by ARO grant W911NF1920041, Brazilian agencies CAPES (Gilvania Vilela/POS-DOC-88881.120327/2016-01), FACEPE (APQ-0565-1.05/14), CNPq and UPE (PFA/PROGRAD/UPE 04/2017). M.G.H., J.L.M. and N.M.N. were supported by Spanish MICIN grants MAT2017-87134- C02 and MAT2017-84496-R.
\end{acknowledgement}
	
The Supporting Information is available free of charge on the ACS Publications website at DOI: XXXX. 	The SI contains results of measurements on additional devices of NbN and NbN/GdN bilayers; and further details of the proposed theoretical model for the mechanism.

\providecommand{\latin}[1]{#1}
\makeatletter
\providecommand{\doi}
  {\begingroup\let\do\@makeother\dospecials
  \catcode`\{=1 \catcode`\}=2 \doi@aux}
\providecommand{\doi@aux}[1]{\endgroup\texttt{#1}}
\makeatother
\providecommand*\mcitethebibliography{\thebibliography}
\csname @ifundefined\endcsname{endmcitethebibliography}
  {\let\endmcitethebibliography\endthebibliography}{}

\end{document}